\documentclass[12pt]{article}
\usepackage[margin=1in]{geometry}
\usepackage{bm}
\usepackage{graphicx}
\usepackage{amssymb,amsmath,amsthm,mathrsfs}
\usepackage{titlesec}
\usepackage{epstopdf}
\usepackage{algorithm}
\usepackage{amsfonts,delarray}
\usepackage{rotating,color}
\usepackage{subfigure}
\usepackage{natbib,multirow}
\usepackage{titlesec,textcase}
\usepackage{longtable}
\usepackage{bbm}
\usepackage{rotating}
\usepackage{authblk}
\usepackage{times}
\usepackage{graphicx}
\usepackage{subfigure}
\usepackage{hyperref}
\usepackage{dblfloatfix}
\usepackage[flushleft]{threeparttable}
\usepackage{arydshln}

\titleformat{\section}{\bf\large\center\uppercase}{\thesection.}{1em}{}

\def\balpha{\boldsymbol\alpha}
\def\btheta{\boldsymbol\theta}
\def\bomega{\boldsymbol\omega}
\def\htheta{\hat\theta}

\def\bone{\boldsymbol 1}
\def\ba{\boldsymbol a}
\def\bA{\boldsymbol A}
\def\bb{\boldsymbol b}

\def\bD{\boldsymbol D}
\def\bL{\boldsymbol L}

\def\bu{\boldsymbol u}
\def\bH{\boldsymbol H}
\def\bV{\boldsymbol V}
\def\bW{\boldsymbol W}
\def\diag{\text{diag}}

\newcommand{\norm}[1]{\left\lVert#1\right\rVert}
\newcommand{\var}{\mbox{Var}}
\newcommand{\cov}{\mbox{Cov}}

\DeclareMathOperator*{\argmin}{arg\,min}

\DeclareMathOperator*{\median}{median}
\definecolor{rp}{RGB}{83,54,106}
\begin{document}

\title{How Many Communities Are There?
\thanks{Diego Franco Saldana (Email: diego@stat.columbia.edu) and Yang Feng (Email: yangfeng@stat.columbia.edu),  Department of Statistics, Columbia University, New York, NY 10027. Yi Yu (Email: y.yu@statslab.cam.ac.uk),  Statistical Laboratory, University of Cambridge, U.K. CB30WB.}
}
\author{Diego Franco Saldana, Yi Yu, and Yang Feng}

\date{}
\maketitle

\begin{abstract}
Stochastic blockmodels and variants thereof are among the most widely used approaches to community detection for social networks and relational data. A stochastic blockmodel partitions the nodes of a network into disjoint sets, called communities. The approach is inherently related to clustering with mixture models; and raises a similar model selection problem for the number of communities. The Bayesian information criterion (BIC) is a popular solution, however, for stochastic blockmodels, the conditional independence assumption given the communities of the endpoints among different edges is usually violated in practice. In this regard, we propose composite likelihood BIC (CL-BIC) to select the number of communities, and we show it is robust against possible misspecifications in the underlying stochastic blockmodel assumptions. We derive the requisite methodology and illustrate the approach using both simulated and real data. Supplementary materials containing the relevant computer code are available online.

\end{abstract}

\section{Introduction}

Enormous network datasets are being generated and analyzed along with an increasing interest from researchers in studying the underlying structures of a complex networked world. The potential benefits span traditional scientific fields such as epidemiology and physics, but also emerging industries, especially large-scale internet companies. Among a variety of interesting problems arising with network data, in this paper, we focus on community detection in undirected networks $G := (V, E)$, where $V$ and $E$ are the sets of nodes and edges, respectively. In this framework, the community detection problem can be formulated as finding the true disjoint partition of $V = V_1 \sqcup \cdots \sqcup V_K$, where $K$ is the number of communities. Although it is difficult to give a rigorous definition, communities are often regarded as tightly-knit groups of nodes which are loosely connected between themselves.

The community detection problem has close connections with graph partitioning, which could be traced back to Euler, while it has its own characteristics due to the concrete physical meanings from the underlying dataset \citep{NewmanGirvan2004}. Over the last decade, there has been a considerable amount of work on it, including minimizing ratio cut \citep{WeiCheng1989}, minimizing normalized cut \citep{ShiMalik2000}, maximizing modularity \citep{NewmanGirvan2004}, hierarchical clustering \citep{Newman2004} and edge-removal methods \citep{NewmanGirvan2004}, to name a few. Among all the progress made by peer researchers, spectral clustering \citep{DonathEtal1973} based on stochastic blockmodels \citep{HollandEtal1983} debuted and soon gained a majority of attention. We refer the interested readers to \citet{SpielmatTeng1996} and \citet{GoldenbergEtal2010} as comprehensive reviews on the history of spectral clustering and stochastic blockmodels, respectively.

Compared to the amount of work on spectral clustering or stochastic blockmodels, to the best of our knowledge, there is little work on the selection of the community number $K$. In most of the previously mentioned community detection methods, the number of communities is generally input as a pre-specified quantity. For the literature addressing the problem of selecting $K$, besides the block-wise edge splitting method of \cite{ChenLei2014}, a common practice is to use BIC-type criteria \citep{AiroldiEtal2008,DaudinEtal2008} or a variational Bayes approach \citep{LatoucheEtal2012,HunterEtal2012}. An inherently related problem is that of selecting the number of components in mixture models, where the birth-and-death point process of \cite{Stephens2000} and the allocation sampler of \cite{NobileFearnside2007} provide two fully Bayesian approaches in the case where $K$ is finite but unknown. Based on the allocation sampler, \cite{McCaidEtal2013} propose an efficient Bayesian clustering algorithm which directly estimates the number of communities in stochastic blockmodels, and which exhibits similar results to the variational Bayes approach of \cite{LatoucheEtal2012}. Nonparametric Bayesian methods based on Dirichlet process mixtures \citep{Ferguson1973} have also been used to estimate the number of components in this finite but unknown $K$ setting \citep{Fearnhead2004}, although the inconsistency of this approach has been recently shown by \cite{MillerHarrison2014}. This community or mixture component number $K$, as a vital part of model selection procedures, highly depends on the model assumptions. For instance, the famous stochastic blockmodel has undesirable restrictive assumptions in the form of independent Bernoulli observations when the community assignments are known.

In this paper, we study the community number selection problem with robustness consideration against model misspecification in the stochastic blockmodel and its variants. Our motivation is that, the conditional independence assumption among edges, when the communities of their endpoints are given, is usually violated in real applications. In addition, we do not restrict our interest only to exchangeable graphs. Using  techniques from the composite likelihood paradigm \citep{Lindsay1988}, we develop a composite likelihood BIC (CL-BIC) approach \citep{GaoSong2010} for selecting the community number in the situation where assumed independencies in the stochastic blockmodel and other exchangeable graph models do not hold. The procedure is tested on simulated and real data, and is shown to outperform two competitors -- traditional BIC and the variational Bayes criterion of \cite{LatoucheEtal2012}, in terms of model selection consistency.

The rest of the paper is organized as follows. The background for stochastic blockmodels and spectral clustering is introduced in Section \ref{background}, and the proposed CL-BIC methodology  is developed in Section \ref{methodology}. In Section \ref{numerical}, several simulation examples as well as two real data sets are analyzed. The paper is concluded with a short discussion in Section \ref{discussion}.

\section{Background}\label{background}

First, we would like to introduce some notation. For an $N$-node undirected, simple, and connected network $G$, its symmetric adjacency matrix $\bA$ is defined as $A_{ij} := 1$ if $(i, j)$ is an element in $E$, and $A_{ij} := 0$ otherwise. The diagonal $\{A_{ii}\}_{i=1}^{N}$ is fixed to zero (i.e., self-edges are not allowed). Moreover, $\bD$ and $\bL$  denote the degree matrix and Laplacian matrix, respectively. Here, ${D}_{ii} := d_i$, and ${D}_{ij} := 0$ for $i \neq j$, where $d_i$ is the degree of node $i$, i.e., the number of edges with endpoint node $i$; and $\bL :=\bD^{-1/2}\bA\bD^{-1/2}$. As isolated nodes are discarded, $\bD^{-1/2}$ is well-defined.

\subsection{Stochastic Blockmodels}

\subsubsection{Standard Stochastic Blockmodel}

Stochastic blockmodels were first introduced in \citet{HollandEtal1983}. They posit independent Bernoulli random variables $\{A_{ij}\}_{1 \leq i < j \leq N}$ with success probabilities $\{P_{ij}\}$ which depend on the communities of their endpoints $i$ and $j$. Consequently, all edges are conditionally independent given the corresponding communities. Moreover, each node is associated with one and only one community, with label $Z_i$, where $Z_i \in \{1, \ldots, K\}$. Following \citet{RoheEtal2011} and \citet{ChoiEtal2012}, throughout this paper we assume each $Z_i$ is fixed and unknown, thus yielding $\mathbb{P}(A_{ij}=1; Z_i = z_i, Z_j = z_j) = \theta_{z_i z_j}$. Treating the node assignments $Z_1,\ldots,Z_N$ as latent random variables is another popular approach in the community detection literature, and various methods including the variational Bayes criterion of \citet{LatoucheEtal2012} and the belief propagation algorithm of \citet{DecelleEtal2012} efficiently approximate the corresponding \emph{observed-data} log-likelihood of the stochastic blockmodel, without having to add $K^N$ multinomial terms accounting for all possible label assignments.

For $\btheta := (\theta_{ab}; 1\le a \le b \le K)'$ and for any fixed community assignment $\boldsymbol{z} \in \{1, \ldots, K\}^N$, the log-likelihood under the standard \emph{Stochastic Blockmodel} (SBM) is given as
\begin{equation}\label{eq-lik}
\ell(\btheta; \bA)  := \sum_{i < j}[A_{ij} \log \theta_{z_iz_j} + (1-A_{ij})\log (1-\theta_{z_iz_j})].
\end{equation}

For the remainder of the paper, denote $N_a$ as the size of community $a$, and $n_{ab}$ as the maximum number of possible edges between communities $a$ and $b$, i.e., $n_{ab} := N_aN_b$ for $a\neq b$, and $n_{aa} := N_a(N_a-1)/2$.  Also, let $m_{ab} := \sum_{i<j}A_{ij}\bone\{z_i = a, z_j = b\}$, and $\htheta_{ab} := m_{ab}/n_{ab}$ be the MLE of $\theta_{ab}$ in \eqref{eq-lik}.

Under this framework, \cite{ChoiEtal2012} showed that the fraction of misclustered nodes converges in probability to zero under maximum likelihood fitting when $K$ is allowed to grow no faster than $\sqrt{N}$. By means of a \emph{regularized} maximum likelihood estimation approach, \cite{RoheEtal2014} further proved that this \emph{weak} convergence can be achieved for $K=\mathcal{O}(N/\log^5 N)$.

\subsubsection{Degree-Corrected Stochastic Blockmodel}

Heteroscedasticity of node degrees within communities is often observed in real-world networks. To tackle this problem, \citet{KarrerNewman2011} proposed the \emph{Degree-Corrected Blockmodel} (DCBM), in which the success probabilities $\{P_{ij}\}$ are also functions of individual effects. To be more precise, the DCBM assumes that $\mathbb{P}(A_{ij}=1; Z_i = z_i, Z_j = z_j) = \omega_i\omega_j\theta_{z_i z_j}$, where $\boldsymbol{\omega} := (\omega_1, \ldots, \omega_n)'$ are individual effect parameters satisfying the identifiability constraint $\sum_{i}\omega_i\bone\{z_i = a\} = 1$ for each community $1 \leq a \leq K$.

To simplify technical derivations, \citet{KarrerNewman2011} allowed networks to contain both multi-edges and self-edges. Thus, they assumed the random variables $\{A_{ij}\}_{1 \leq i \leq j \leq N}$ to be independent Poisson, with the previously defined success probabilities $\{P_{ij}\}$ of an edge between vertices $i$ and $j$ replaced by the expected number of such edges. Under this framework, and for any fixed community assignment $\boldsymbol{z} \in \{1, \ldots, K\}^N$, \citet{KarrerNewman2011} arrived at the log-likelihood $\ell(\boldsymbol{\theta, \omega}; \bA)$ of observing the adjacency matrix $\bA = (A_{ij})$ under the DCBM,
\begin{align}\label{eq-lik-dcbm}
\ell(\boldsymbol{\theta, \omega}; \bA) := 2\sum_i d_i\log \omega_i + \sum_{a, b}(m_{ab}\log \theta_{ab} - \theta_{ab}).
\end{align}
After allowing for the identifiability constraint on $\boldsymbol{\omega}$, the MLEs of the parameters $\theta_{ab}$ and $\omega_i$ are given by $\hat{\theta}_{ab} := m_{ab}$ and $\hat{\omega}_i := d_i/\sum_{j: z_j = z_i}d_j$, respectively.

As mentioned in \citet{ZhaoEtal2012}, there is no practical difference in performance between the log-likelihood \eqref{eq-lik-dcbm} and its slightly more elaborate version based on the true Bernoulli observations. The reason is that the Bernoulli distribution with a small mean is well approximated by the Poisson distribution, and the sparser the network is, the better the approximation works \citep{PerryWolfe2012}.

\subsubsection{Mixed Membership Stochastic Blockmodel}

As a methodological extension in which nodes are allowed to belong to more than one community, \cite{AiroldiEtal2008} proposed the \emph{Mixed Membership Stochastic Blockmodel} (MMB) for directed relational data $\{A_{ij}\}_{1 \leq i,j \leq N}$. For instance, when a social actor interacts with its different neighbors, an array of different social contexts may be taking place and thus the actor may be taking on different latent roles.

The model assumes the observed network is generated according to node-specific distributions of community membership and edge-specific indicator vectors denoting membership in one of the $K$ communities. More specifically, each vertex $i$ is associated with a randomly drawn vector $\vec{\boldsymbol \pi}_i$, with $\pi_{ia}$ denoting the probability of node $i$ belonging to community $a$. Additionally, let the indicator vector $\vec{ \boldsymbol z}_{i \rightarrow j}$ denote the community membership of node $i$ when he sends a message to node $j$, and $\vec{\boldsymbol z}_{i \leftarrow j}$ denote the community membership of node $j$ when he receives a message from node $i$. If, in order to account for the asymmetric interactions, we denote by $\btheta := (\theta_{ab})$ the $K \times K$ matrix where $\theta_{ab}$ represents the probability of having an edge from a social actor in community $a$ to a social actor in community $b$, the MMB posits that the $\{A_{ij}\}_{1 \leq i,j \leq N}$ are drawn from the following generative process.

\begin{itemize}
\item For each node $i \in V$:
\begin{itemize}
\item Draw a $K$ dimensional mixed membership vector $\vec{\boldsymbol \pi}_i \sim \text{Dirichlet}(\boldsymbol \alpha)$, with the vector $\boldsymbol \alpha=(\alpha_1,\ldots,\alpha_K)'$ being a hyper-parameter.
\end{itemize}
\item For each possible edge variable ${A_{ij}}$:
\begin{itemize}
\item Draw membership indicator vector for the initiator $\vec{\boldsymbol z}_{i \rightarrow j} \sim \text{Multinomial}(\vec{\boldsymbol \pi}_i)$.
\item Draw membership indicator vector for the receiver $\vec{\boldsymbol z}_{i \leftarrow j} \sim \text{Multinomial}(\vec{\boldsymbol \pi}_j)$.
\item Sample the interaction $A_{ij} \sim \text{Bernoulli}(\vec{ \boldsymbol z}_{i \rightarrow j}' \btheta \vec{\boldsymbol z}_{i \leftarrow j})$.
\end{itemize}
\end{itemize}

Upon defining the set of mixed membership vectors $\boldsymbol{\Pi}:= \{ \vec{\boldsymbol \pi}_i : i \in V \}$ and the sets of membership indicator vectors $\boldsymbol{Z}_{\rightarrow}:=\{ \vec{\boldsymbol z}_{i \rightarrow j} : i,j \in V \}$ and $\boldsymbol{Z}_{\leftarrow}:=\{ \vec{\boldsymbol z}_{i \leftarrow j} : i,j \in V \}$, following \cite{AiroldiEtal2008}, we obtain the complete data log-likelihood of the \emph{hyper-parameters} $\{ \boldsymbol{\theta}, \boldsymbol{\alpha} \}$ as
\begin{equation}\label{eq-lik-mmb}
\begin{aligned}
\ell(\boldsymbol{\theta, \alpha}; \bA, \boldsymbol{\Pi}, \boldsymbol{Z}_{\rightarrow}, \boldsymbol{Z}_{\leftarrow}) & := \sum_{i,j}[A_{ij} \log (\vec{ \boldsymbol z}_{i \rightarrow j}' \btheta \vec{\boldsymbol z}_{i \leftarrow j}) + (1-A_{ij})\log (1-\vec{ \boldsymbol z}_{i \rightarrow j}' \btheta \vec{\boldsymbol z}_{i \leftarrow j})] \\
 & \mkern-10mu + N \Big( \log \Gamma (\sum_{a} \alpha_a) - \sum_{a} \log \Gamma(\alpha_a) \Big) + \sum_{i} \sum_{a} ( \alpha_a - 1)\log \pi_{ia} + \text{const},
\end{aligned}
\end{equation}
where $\bA$ corresponds to the observed data and $\{ \boldsymbol{\Pi}, \boldsymbol{Z}_{\rightarrow}, \boldsymbol{Z}_{\leftarrow} \}$ are the latent variables.

In order to carry out posterior inference of the latent variables given the observations $\bA$, \cite{AiroldiEtal2008} proposed an efficient coordinate ascent algorithm based on a variational approximation to the true posterior. Therefore, one can compute expected posterior mixed membership vectors and posterior membership indicator vectors. We refer interested readers to Section 3 in \cite{AiroldiEtal2008} for further details.

Consequently, following the same profile likelihood approach, for any fixed set $\{ \boldsymbol{\Pi}, \boldsymbol{Z}_{\rightarrow}, \boldsymbol{Z}_{\leftarrow} \}$, the MLE of $\theta_{ab}$ is given by
\begin{equation}
\htheta_{ab} := \sum_{i,j} A_{ij} \cdot \vec{\boldsymbol z}_{i \rightarrow j, a} \vec{\boldsymbol z}_{i \leftarrow j, b} \bigg/ \sum_{i,j} \vec{\boldsymbol z}_{i \rightarrow j, a} \vec{\boldsymbol z}_{i \leftarrow j, b}.
\end{equation}
As the MLE of $\alpha_a$ does not admit a closed form, \cite{Minka2000} proposed an efficient Newton-Raphson procedure for obtaining parameter estimates in Dirichlet models, where the gradient and Hessian matrix of the complete data log-likelihood (\ref{eq-lik-mmb}) with respect to $\boldsymbol{\alpha}$ are
\begin{equation}\label{newton-mmb}
\begin{aligned}
\frac{\partial \ell(\boldsymbol{\theta, \alpha}; \bA)}{\partial \alpha_a} = & \,\, N \Big( \Psi(\sum_{a} \alpha_a) - \Psi(\alpha_a) \Big) + \sum_{i} \sum_{a} \log \pi_{ia} \\
\frac{\partial^2 \ell(\boldsymbol{\theta, \alpha}; \bA)}{\partial \alpha_a \partial \alpha_b} = & \,\, N \Big( \Psi'(\sum_{a} \alpha_a) - \Psi'(\alpha_a) \bone\{ a=b \} \Big),
\end{aligned}
\end{equation}
and $\Psi$ is known as the digamma function (i.e., the logarithmic derivative of the gamma function).

\subsection{Spectral Clustering and SCORE}

Although there is a parametric framework for the standard stochastic blockmodel, considering the computational burden, it is intractable to directly estimate both parameters $\btheta$ and $\boldsymbol{z}$  based on exact maximization of the log-likelihood (\ref{eq-lik}). Researchers have instead resorted to spectral clustering as a computationally feasible algorithm. For comprehensive reviews, we refer interested readers to \citet{Luxburg2007} and \citet{RoheEtal2011}, in which the authors proved the consistency of spectral clustering in the standard stochastic blockmodel under proper conditions imposed on the density of the network and the eigen-structure of the Laplacian matrix. The algorithm finds the eigenvectors $\boldsymbol{u_1}, \ldots, \boldsymbol{u_K}$ associated with the $K$ eigenvalues of $\bL$ that are largest in magnitude, forming an $N \times K$ matrix $\boldsymbol{U} := (\boldsymbol{u_1}, \ldots, \boldsymbol{u_K})$, and then applies the $K$-means algorithm to the rows of $\boldsymbol{U}$.

Similarly, \citet{Jin2015} proposed a variant of spectral clustering for the DCBM, called Spectral Clustering On Ratios-of-Eigenvectors (SCORE). Instead of using the Laplacian matrix $\bL$, SCORE collects the eigenvectors $\boldsymbol{v_1}, \ldots, \boldsymbol{v_K}$ associated with the $K$ eigenvalues of $\bA$ that are largest in magnitude, and then forms the $N \times K$ matrix $\boldsymbol{V} := (\boldsymbol{1}, \boldsymbol{v_2}/\boldsymbol{v_1}, \ldots, \boldsymbol{v_K}/\boldsymbol{v_1})$, where the division operator is taken entry-wise, i.e., for vectors $\ba,\bb \in \mathbb{R}^n$, with $\Pi_{\ell = 1}^n b_{\ell} \neq 0$, $\ba/\bb := (a_1/b_1, \ldots, a_n/b_n)'$. SCORE then applies the $K$-means algorithm to the rows of $\boldsymbol{V}$. The corresponding consistency results for the DCBM are also provided in \cite{Jin2015}.

\section{Model Selection for the Number of Communities}\label{methodology}

\subsection{Motivation}

In much of the previous work, e.g. \citet{AiroldiEtal2008}, \citet{DaudinEtal2008} and \citet{HandcockEtal2007}, researchers have used a BIC-penalized version of the log-likelihood \eqref{eq-lik} to choose the community number $K$. However, we are aware of the possible misspecifications in the underlying stochastic blockmodel assumptions and in the loss of precision from the computational relaxation brought in by spectral clustering.

Firstly, in network data, edges are not necessarily independent if only the communities of their endpoints are given. For instance, if two different edges $A_{ij}$ and $A_{il}$ have mutual endpoint $i$, it is highly likely that they are dependent even given the community labels of their endpoints. This misspecification problem exists in both the standard stochastic blockmodel and its variants, such as DCBM \citep{KarrerNewman2011} and  MMB \citep{AiroldiEtal2008}. Secondly, as  previously  mentioned, spectral clustering is a feasible relaxation, but the loss of precision is inevitable. Several examples of this can be found in \citet{GuatteryMiller1998}. Whence, we resort to introducing CL-BIC with the concern of robustness against misspecifications in the underlying stochastic blockmodel.

We would like to emphasize that CL-BIC is not a new community detection method. Instead, under the SBM, DCBM, or MMB assumptions, it can be combined with existing community detection methods to choose the true community number.

\subsection{Composite Likelihood Inference}

The CL-BIC approach extends the concepts and theory of conventional BIC on likelihoods to the composite likelihood paradigm \citep{Lindsay1988,VarinEtal2011}. Composite likelihood aims at a relaxation of the computational complexity of statistical inference based on exact likelihoods. For instance, when the dependence structure for relational data is too complicated to implement, a working independence assumption can effectively recover some properties of the usual maximum likelihood estimators \citep{CoxReid2004,VarinEtal2011}. However, under this misspecification framework, the asymptotic variance of the resulting estimators is usually underestimated as the Fisher information. Composite marginal likelihoods (also known as \textit{independence likelihoods}) have the same formula as conventional likelihoods in terms of being a product of marginal densities \citep{Varin2008}, while statistical inference based on them can capture this loss of variance. Consequently, to pursue the ``true'' model, CL-BIC penalizes the number of parameters more than what BIC does for dependent relational data.

Before going into details, we would like to give the rationale of using stochastic blockmodels under a misspecification framework. In order to estimate the true joint density $g$ of  $\{A_{ij}\}_{1 \leq i < j \leq N}$, we consider the stochastic blockmodel family $\mathscr{P} = \{ p_{\btheta} : \btheta \in \Theta \}$, where $\Theta = [0,1]^{K(K+1)/2}$ for the standard stochastic blockmodel, and $\Theta = [0,1]^{K(K+1)/2 + N}$ for DCBM. The true joint density $g$ may or may not belong to $\mathscr{P}$, which is a parametric family imposing independence among the $\{A_{ij}\}_{i<j}$ when only the communities of the endpoints are given.

Due to the difficulty in specifying the full, highly structured ${N \choose 2}$-dimensional density $g$, while having access to the univariate densities $p_{ij}(\cdot;\btheta)$ of $A_{ij}$ under the blockmodel family $\mathscr{P}$, the composite marginal likelihood paradigm compounds the first-order log-likelihood contributions to form the composite log-likelihood
\begin{equation}\label{cl-generic}
\text{cl}(\btheta; \bA) := \sum_{i < j} \log \, p_{ij}(A_{ij};\btheta),
\end{equation}
where $\text{cl}(\cdot; \bA)$ corresponds to \eqref{eq-lik} under the standard stochastic blockmodel, and corresponds to \eqref{eq-lik-dcbm} in the DCBM framework. Since each component of $\text{cl}(\btheta; \bA)$ in \eqref{cl-generic} is a valid log-likelihood object, the \textit{composite score estimating equation} $\nabla_{\btheta} \, \text{cl}(\btheta; \bA)=0$ is unbiased under usual regularity conditions. The associated Composite Likelihood Estimator (CLE) $\boldsymbol{\hat{\theta}}_{\text{C}}$, defined as the solution to $\nabla_{\btheta} \, \text{cl}(\btheta; \bA)=0$, suggests a natural estimator of the form $\hat{p}=p_{\boldsymbol{\hat{\theta}}_{\text{C}}}$ to minimize the expected composite Kullback--Leibler divergence \citep{VarinVidoni2005} between the assumed blockmodel $p_{\btheta}$ and the true, but unknown, joint density $g$,
\begin{equation*}
\Delta_{\text{C}}(g,p;\btheta) := \sum_{i < j} \mathbb{E}_g (\log \, g(\{A_{ij}\}_{i<j} \in \mathscr{A}_{ij}) - \log \, p_{ij}(A_{ij};\btheta)),
\end{equation*}
where $\{ \mathscr{A}_{ij} \}_{i<j}$ denotes the corresponding set of marginal events.

In terms of the asymptotic properties of the CLE, following the discussion in \cite{CoxReid2004}, it is important to distinguish whether the available data consist of many independent replicates from a common distribution function or form a few individually large sequences. While, in the first scenario, consistency and asymptotic normality of the corresponding $\boldsymbol{\hat{\theta}}_{\text{C}}$ hold under some regularity conditions from the classical theory of estimating equations \citep{VarinEtal2011}, some difficulties arise in the second one, which includes our observations $\{A_{ij}\}_{i<j}$. Indeed, as argued in \cite{CoxReid2004}, if there is too much internal correlation present among the individual components of the composite score $\nabla_{\btheta} \, \text{cl}(\btheta; \bA)$, the estimator $\boldsymbol{\hat{\theta}}_{\text{C}}$ will not be consistent. The CLE will retain good properties as long as the data are not too highly correlated, which is the case for spatial data with exponential correlation decay. Under this setting, \cite{HeagertyLele1998} proved consistency and asymptotic normality of $\boldsymbol{\hat{\theta}}_{\text{C}}$ in a scenario where the data are not sampled independently from a study population. Under more general settings, consistency results are expected upon using limit theorems and parametric estimation for fields \citep[e.g.][]{Guyon1995}; however, applying the corresponding results requires a properly defined distance on networks and $\alpha$-mixing conditions based on such distance.

\subsection{Composite Likelihood BIC}

Taking into account the measure of model complexity in the context of composite marginal likelihoods \citep{VarinVidoni2005}, we define the following criterion for selecting the community number $K$:
\begin{equation}\label{cl-bic-pop}
\text{CL-BIC}_k := -2\,\text{cl}(\boldsymbol{\hat{\theta}}_{\text{C}}; \bA) + d_k^* \log \left( N(N-1)/2 \right),
\end{equation}
where $k$ is the number of communities under consideration in the current model used as model index, $d^*_k := \text{trace}(\bH_k^{-1}\bV_k)$, $\bH_k := E_{\btheta}(-\nabla_{\btheta}^2 \, \text{cl}(\btheta; \bA))$ and $\bV_k := \text{Var}_{\btheta}(\nabla_{\btheta} \, \text{cl}(\btheta; \bA))$. Then the resulting estimator for the community number is
$$
\hat{k}_{CL-BIC} := \argmin_{k} \text{CL-BIC}_k.
$$

Note that the CLE is a function of $k$, since a different model index yields a different estimator $\boldsymbol{\hat{\theta}}_{\text{C}}:= \boldsymbol{\hat{\theta}}_{\text{C}}(k)$.  Assuming independent and identically distributed data replicates, which lead to consistent and asymptotically normally distributed estimators $\boldsymbol{\hat{\theta}}_{\text{C}}$, \cite{GaoSong2010} established the model selection consistency of a similar composite likelihood BIC approach for high-dimensional parametric models. While allowing for the number of potential model parameters to increase to infinity, their consistency result only holds when the true model sparsity is bounded by a universal constant.

Even though, under a misspecification framework for the blockmodel family $\mathscr{P}$, the observed data $\{A_{ij}\}_{i<j}$ do not form independent replicates from a common population, we anticipate the CL-BIC criterion \eqref{cl-bic-pop} to be consistent in selecting the true community number $K$, at least when the correlation among the $\{A_{ij}\}_{i<j}$ is not severe and the estimators $\boldsymbol{\hat{\theta}}_{\text{C}}$ are consistent and asymptotically normal, as in \cite{HeagertyLele1998}. Since all the moment conditions in the consistency results from \citet{GaoSong2010} hold automatically after noticing the specific forms of the blockmodel composite log-likelihoods \eqref{eq-lik} -- \eqref{eq-lik-mmb}, under a properly defined mixing condition on $\{A_{ij}\}_{i<j}$ \citep{Guyon1995}, and for a bounded community number $K \leq k_0$, we conjecture that $\mathbb{P}\{\hat{k}_{CL-BIC} = K\} \rightarrow 1$ as the number of nodes $N$ in the network grows to infinity. This  theoretical study will be relegated as a future work.

\subsection{Formulae}

\subsubsection{Standard Stochastic Blockmodel}
Following our discussions in the previous section, we treat \eqref{eq-lik} as the composite marginal likelihood, under the working independence assumption that, given the community labels of their endpoints, the Bernoulli random variables $\{A_{ij}\}_{i<j}$ are independent. The first-order partial derivative of $\ell(\btheta; \bA)$ with respect to $\btheta$ is denoted as $\bu(\btheta) = (u(\theta_{ab}); 1\le a \le b \le k)'$, where
\begin{align*}
u(\theta_{ab}) = \sum_{i<j}\left[ \frac{A_{ij}}{\theta_{z_iz_j}} - \frac{1-A_{ij}}{1-\theta_{z_iz_j}} \right] \boldsymbol{I}^{a,b}_{i,j},
\end{align*}
and
\begin{align*}
\boldsymbol{I}^{a, b}_{i, j} = \min\left(\mathbbm{1}\{z_i = a, z_j = b\} + \mathbbm{1}\{z_i = b, z_j = a\}, 1\right).
\end{align*}
Furthermore, the second-order partial derivative of $\ell(\btheta; \bA)$ has the following components,
\begin{align*}
& \hspace{-0.3in} \frac{\partial^2 \ell(\btheta; \bA)}{\partial \theta_{a_1b_1}\partial \theta_{a_2b_2}} = 0, \; \text{ if } (a_1, b_1) \neq (a_2, b_2)
\end{align*}
and
\begin{align*}
& \hspace{0.52in} \frac{\partial^2 \ell(\btheta; \bA)}{\partial \theta_{ab}^2} = -\sum_{i< j} \left[\frac{A_{ij}}{\theta^2_{z_iz_j}} + \frac{1-A_{ij}}{(1-\theta_{z_iz_j})^2}\right]\boldsymbol{I}^{a, b}_{i, j}.
\end{align*}
Define the Hessian matrix $\bH_k(\btheta) = E_{\btheta}(-\partial \bu(\btheta)/\partial \btheta)$, then
\begin{align*}
\bH_k(\btheta) = E_{\btheta}\left(\diag\left\{-\partial^2 \ell(\btheta; \bA)/\partial \theta_{ab}^2; 1\le a \le b \le k\right\}\right).
\end{align*}
Define the variability matrix $\bV_k(\btheta) = \text{Var}_{\btheta}(\bu(\btheta))$ and, following \cite{VarinVidoni2005}, the model complexity $d_{k}^* = \text{trace}[\bH_k(\btheta)^{-1}\bV_k(\btheta)]$. If the underlying model is indeed a correctly specified standard stochastic blockmodel, we have $d_{k}^* = k(k+1)/2$ and CL-BIC reduces to the traditional BIC. Indexed by $1 \le k \le k_0$, the estimated criterion functions for the CL-BIC sequence \eqref{cl-bic-pop} are
\begin{equation}\label{criterion}
\widehat{\text{CL-BIC}_k} = -2\,\text{cl}(\boldsymbol{\hat{\theta}}_{\text{C}}; \bA) + \hat{d}^*_k \log \left( N(N-1)/2 \right),
\end{equation}
where $\boldsymbol{\hat{\theta}}_{\text{C}}$ and $\hat{d}^*_k$ are estimators of $\btheta$ and $d^*_k$, respectively. For a certain $k$, the explicit estimator forms are given below:
\begin{align*}
& \boldsymbol{\hat{H}}_k(\boldsymbol{\hat{\theta}}_{\text{C}}) = \diag\left\{\sum_{i< j} \left[ \frac{A_{ij}}{\htheta^2_{z_iz_j}} + \frac{(1-A_{ij})}{(1-\htheta_{z_iz_j})^2} \right] \boldsymbol{I}^{a,b}_{i,j} \right\}
\end{align*}
and
$\boldsymbol{\hat{V}}_k(\boldsymbol{\hat{\theta}}_{\text{C}}) = \bu(\boldsymbol{\hat{\theta}}_{\text{C}})[\bu(\boldsymbol{\hat{\theta}}_{\text{C}})]^T$.

As noted in \citet{GaoSong2010}, the above naive estimator for $\bV_k(\btheta)$ vanishes when evaluated at the CLE $\boldsymbol{\hat{\theta}}_{\text{C}}$. An alternative proposed in \citet{VarinEtal2011} is to use a jackknife covariance matrix estimator, for the asymptotic covariance matrix of $\boldsymbol{\hat{\theta}}_{\text{C}}$, of the form
\begin{align}\label{eq-jack}
& \var_{\text{jack}}(\boldsymbol{\hat{\theta}}_{\text{C}}) = \frac{N-1}{N} \sum_{l=1}^{N} (\boldsymbol{\hat{\theta}}_{\text{C}}^{(-l)} - \boldsymbol{\hat{\theta}}_{\text{C}})(\boldsymbol{\hat{\theta}}_{\text{C}}^{(-l)} - \boldsymbol{\hat{\theta}}_{\text{C}})^T,
\end{align}
where $\boldsymbol{\hat{\theta}}_{\text{C}}^{(-l)}$ is the composite likelihood estimator of $\btheta$ with the $l$-th vertex deleted. Let $\bA^{(-l)}$ be the $(N-1) \times (N-1)$ matrix obtained after deleting the $l$-th row and column from the original adjacency matrix $\bA$. An explicit form for $\boldsymbol{\hat{\theta}}_{\text{C}}^{(-l)}$ is given by $\htheta_{ab}^{(-l)}  = 1/n_{ab}^{(-l)} \sum_{i<j} A_{ij}^{(-l)} \bone\{z_i = a, z_j = b\}$, with $n_{ab}^{(-l)} = N_a^{(-l)} N_b^{(-l)}$ for $a\neq b$, and $n_{aa}^{(-l)} = N_a^{(-l)}(N_a^{(-l)} - 1)/2$; naturally, $N_a^{(-l)} = N_a - 1$ if $z_l =a$ and $N_a^{(-l)} = N_a$ otherwise.

Since the asymptotic covariance matrix of $\boldsymbol{\hat{\theta}}_{\text{C}}$ is given by the inverse Godambe information matrix, $G_k(\btheta)^{-1} = \bH_k(\btheta)^{-1} \bV_k(\btheta) \bH_k(\btheta)^{-1}$, see \citet{GaoSong2010} and \citet{VarinEtal2011}, an explicit estimator for $d^*_k$ can be obtained by right-multiplying the jackknife covariance matrix estimator (\ref{eq-jack}) by $\bH_k(\btheta)$ to obtain
\begin{eqnarray*}
\hat{d}^*_k & = & \text{trace}\left[\var_{\text{jack}}(\boldsymbol{\hat{\theta}}_{\text{C}}) \boldsymbol{\hat{H}}_k(\boldsymbol{\hat{\theta}}_{\text{C}})\right] \\
            & = & \sum_{1\le a\le b\le k} \bigg\{ \text{Var}_{\text{jack}}(\hat{\theta}_{ab}) \times \sum_{i< j} \Big[\frac{A_{ij}}{\htheta^2_{z_iz_j}} + \frac{1-A_{ij}}{(1-\htheta_{z_iz_j})^2} \Big] \boldsymbol{I}^{a,b}_{i,j} \bigg\}. \\
\end{eqnarray*}

\subsubsection{Degree-Corrected Stochastic Blockmodel}
Similarly, we develop corresponding parallel results for DCBM. The first- and second-order partial derivatives of $\ell(\boldsymbol{\theta}, \boldsymbol{\omega}; \bA)$ with respect to $\boldsymbol{\theta}$ are defined as follows,
\begin{align*}
& \frac{\partial \ell(\boldsymbol{\theta, \omega}; \bA)}{\partial \boldsymbol{\theta}} = \bu(\boldsymbol{\theta}) = (u(\theta_{ab}); 1\le a\le b\le k)', & & \!\!\!\!\!\!\!\! u(\theta_{ab}) = \frac{m_{ab}}{\theta_{ab}} - 1,\\
& \frac{\partial^2\ell(\boldsymbol{\theta, \omega}; \bA)}{\partial \theta_{a_1b_1}\partial \theta_{a_2b_2}} = 0, \mbox{ if } (a_1, b_1)\neq (a_2, b_2), & & \!\!\!\!\!\!\!\! \frac{\partial^2\ell(\boldsymbol{\theta, \omega}; \bA)}{\partial \theta_{ab}^2} = -\frac{m_{ab}}{\theta_{ab}^2},\\
& \boldsymbol{\hat{H}}_k(\boldsymbol{\hat{\theta}}_{\text{C}}) = \diag\left\{\frac{1}{\hat{\theta}_{ab}}; 1\le a\le b\le k\right\}, &
\end{align*}
which yields
\begin{eqnarray*}
\hspace{0.35in} \hat{d}^*_k & = & \sum_{1\le a\le b\le k}\left\{\text{Var}_{\text{jack}}(\hat{\theta}_{ab})/\hat{\theta}_{ab}\right\}.
\end{eqnarray*}

\subsubsection{Mixed Membership Stochastic Blockmodel}
The estimated model complexity for MMB now involves second-order partial derivatives of $\ell(\boldsymbol{\theta, \alpha}; \bA)$ with respect to the hyper-parameters $\btheta$ and $\balpha$. Upon noticing the form of the first term of the complete data log-likelihood (\ref{eq-lik-mmb}), and recalling the Hessian matrix with respect to $\balpha$ detailed in (\ref{newton-mmb}), it is easy to see that $\boldsymbol{\hat{H}}_k(\boldsymbol{\hat{\theta}}_{\text{C}},\boldsymbol{\hat{\alpha}}_{\text{C}})$ is a block matrix of the form

\begin{equation*}
\boldsymbol{\hat{H}}_k(\boldsymbol{\hat{\theta}}_{\text{C}},\boldsymbol{\hat{\alpha}}_{\text{C}}) = \left(
\begin{array}{c:c}
\boldsymbol{\hat{H}}_k(\boldsymbol{\hat{\theta}}_{\text{C}}) & \boldsymbol{0} \\ \hdashline
\boldsymbol{0} & \boldsymbol{\hat{H}}_k(\boldsymbol{\hat{\alpha}}_{\text{C}})
\end{array}
\right),
\end{equation*}
where $\boldsymbol{\hat{H}}_k(\boldsymbol{\hat{\theta}}_{\text{C}})$ is a $k^2 \times k^2$ diagonal matrix given by
\begin{equation*}
\boldsymbol{\hat{H}}_k(\boldsymbol{\hat{\theta}}_{\text{C}}) = \diag\left\{\sum_{i,j} \left[ \frac{A_{ij}}{\htheta^2_{z_{i \rightarrow j},z_{i \leftarrow j}}} + \frac{(1-A_{ij})}{(1-\htheta_{z_{i \rightarrow j},z_{i \leftarrow j}})^2} \right] \bone\{z_{i \rightarrow j} = a, z_{i \leftarrow j} = b\} \right\},
\end{equation*}
and $\boldsymbol{\hat{H}}_k(\boldsymbol{\hat{\alpha}}_{\text{C}}) = (\boldsymbol{\hat{H}}_k(\boldsymbol{\hat{\alpha}}_{\text{C}})_{ab})$ is a $k \times k$ matrix with entries
\begin{equation*}
\boldsymbol{\hat{H}}_k(\boldsymbol{\hat{\alpha}}_{\text{C}})_{ab} = N \Big(  \Psi'(\hat{\alpha}_a) \bone\{ a=b \} - \Psi'(\sum_{a} \hat{\alpha}_a) \Big).
\end{equation*}
In a slight abuse of notation, we denote by $z_{i \rightarrow j}$ above the label assignment corresponding to node $i$ when he sends a message to node $j$, and similarly for $z_{i \leftarrow j}$. The estimated model complexity is thus $\hat{d}^*_k = \text{trace}[ \var_{\text{jack}}(\boldsymbol{\hat{\theta}}_{\text{C}}, \boldsymbol{\hat{\alpha}}_{\text{C}}) \boldsymbol{\hat{H}}_k(\boldsymbol{\hat{\theta}}_{\text{C}}, \boldsymbol{\hat{\alpha}}_{\text{C}}) ]$, where the jackknife matrix $\var_{\text{jack}}(\boldsymbol{\hat{\theta}}_{\text{C}}, \boldsymbol{\hat{\alpha}}_{\text{C}})$, assuming a similar form as in (\ref{eq-jack}) with $\boldsymbol{\hat{\theta}}_{\text{C}}^{(-l)}$ and $\boldsymbol{\hat{\alpha}}_{\text{C}}^{(-l)}$ estimated as explained in Section \ref{background}, provides the corresponding asymptotic covariance matrix estimator of the CLE $(\boldsymbol{\hat{\theta}}_{\text{C}}, \boldsymbol{\hat{\alpha}}_{\text{C}})$.

We would like to remark that our CL-BIC approach for selecting the community number $K$ extends beyond the realm of stochastic blockmodels. Indeed, both the latent space cluster model of \cite{HandcockEtal2007} and the local dependence model of \cite{SchweinbergerHandcock2015}, as well as any other (composite) likelihood-based approach which requires to select a value of $K$ can employ our proposed CL-BIC methodology for selecting the number of communities. We leave the details of this further investigation for future research.

\section{Experiments}\label{numerical}

In this section, we show the advantages of the CL-BIC approach over the traditional BIC as well as the variational Bayes approach in selecting the true number of communities via simulations and two real datasets.

\subsection{Simulations}

For simplicity of the presentation, we consider only the SBM and the DCBM in our simulations. For each setting, we relax the assumption that the $A_{ij}$'s are conditionally independent given the labels $(Z_i = z_i, Z_j = z_j)$, varying both the dependence structure of the adjacency matrix $\bA \in \mathbb{R}^{N \times N}$ and the value of the parameters $(\btheta, \bomega)$. The models introduced are correlation-contaminated stochastic blockmodels, i.e., we bring different types of correlation into the stochastic blockmodels, both standard and degree-corrected, mimicking real-world networks.

All of our simulated adjacency matrices have independent rows. That is, the binary variables $A_{ik}$ and $A_{jl}$ are independent, whenever $i \neq j$, given the corresponding community labels of their endpoints. However, for a fixed node $i \in V$, correlation does exist across different columns in the binary variables $A_{ij}$ and $A_{il}$. For the standard stochastic blockmodel, correlated binary random variables are generated, following the approach in \citet{LeischEtal1998}, by thresholding a multivariate Gaussian vector with correlation matrix $\boldsymbol{\Sigma}$ satisfying $\Sigma_{jl} = \rho_{jl}$. Specifically, for any choice of $|\rho_{jl}| \leq 1$, we simulate correlated variables $A_{ij}$ and $A_{il}$ such that $\cov (A_{ij},A_{il}) = L(-\mu_j,-\mu_l,\rho_{jl}) - \theta_{z_iz_j} \theta_{z_iz_l}$. Here, following \citet{LeischEtal1998}, we have $L(-\mu_j,-\mu_l,\rho_{jl})=\mathbb{P}(W_j \geq -\mu_j,W_l \geq -\mu_l)$, $\mu_j = \Phi^{-1}(\theta_{z_iz_j})$ and $\mu_l = \Phi^{-1}(\theta_{z_iz_l})$, where $(W_j,W_l)$ is standard bivariate normal with correlation $\rho_{jl}$. Correlated Bernoulli variables for the degree-corrected blockmodel are generated in a similar fashion.

In each experiment, carried over $200$ randomly generated  adjacency matrices, we record the proportion of times the chosen number of communities for each of the different criteria for selecting $K$ agrees with the truth. Apart from CL-BIC and BIC, we also consider the Integrated Likelihood Variational Bayes (VB) approach of \citet{LatoucheEtal2012}.  To estimate the true community number, their method selects the candidate value $k$ which maximizes a variational Bayes approximation to the observed-data log-likelihood.

We restrict attention to candidate values for the true $K$ in the range $k \in \{ 1,\ldots,18 \}$, both in simulations and the real data analysis section. For Simulations 1 -- 3, spectral clustering is used to obtain the community labels for each candidate $k$, whereas in the DCBM setting of Simulation 4, the SCORE algorithm is employed. Additionally, among the incorrectly selected community number trials, we calculate the median deviation between the selected community number and the true $K=4$, as well as its robust standard deviation.

\setcounter{dbltopnumber}{2}

\setlength{\tabcolsep}{2.6pt}

\begin{table*}[t]
\small
\begin{threeparttable}
\caption{\small Comparison of CL-BIC and BIC over $200$ repetitions from Simulation 1, where \emph{Eq} and \emph{Dec} indicate equally correlated and exponential decaying cases, respectively. Both correlation of multivariate Gaussian random variables ($\rho$ MVN) and the corresponding \emph{maximum} correlation between Bernoulli variables ($\rho$ Ber.) are presented.}\label{t-1}
\begin{center}
\begin{small}
\begin{tabular}[h]{r @{.} l c c c c c c c c r @{.} l c c c c c c c}
\hline
\hline
\multicolumn{3}{c}{CORR} & & \multicolumn{2}{c}{PROP} & & \multicolumn{2}{c}{MEDIAN DEV} & & \multicolumn{3}{c}{CORR} & & \multicolumn{2}{c}{PROP} & & \multicolumn{2}{c}{MEDIAN DEV}\\ \cline{1-3} \cline{5-6} \cline{8-9} \cline{11-13} \cline{15-16} \cline{18-19}
\multicolumn{2}{c}{$\rho$ MVN} & $\rho$ Ber. & & CL-BIC & BIC & & CL-BIC & BIC & & \multicolumn{2}{c}{$\rho$ MVN} & $\rho$ Ber. & & CL-BIC & BIC & & CL-BIC & BIC \\
\hline
 0&10           & 0.06 & & 1.00 & 0.40 & & 0.0(0.0) & 2.0(1.5) & & 0&40 & 0.25 & & 1.00 & 0.35 & & 0.0(0.0) & 2.0(1.5)\\
 0&15 \emph{Eq} & 0.09 & & 0.92 & 0.14 & & 1.0(0.0) & 3.0(2.2) & & 0&50 \emph{Dec} & 0.32 & & 1.00 & 0.21 & & 0.0(0.0) & 2.0(1.5)\\
 0&20           & 0.12 & & 0.81 & 0.03 & & 1.0(0.4) & 5.0(3.0) & & 0&60 & 0.40 & & 0.99 & 0.12 & & 1.0(0.0) & 3.0(1.5)\\
\hline
\end{tabular}
\begin{tablenotes}[para,flushleft]
NOTE: CORR, correlation; PROP, proportion; MEDIAN DEV, median deviation. In the MEDIAN DEV columns, results are in the form of median (robust standard deviation).
\end{tablenotes}
\end{small}
\end{center}
\end{threeparttable}
\vspace{-0.1in}
\end{table*}

\setlength{\tabcolsep}{2.4pt}

\begin{table*}[t]
\caption{\small Comparison of CL-BIC and BIC over $200$ repetitions from Simulation 2, where \emph{Ind} indicates $\rho_{jl}=0$ for $j \neq l$. For simplicity, we omit the correlation between the corresponding Bernoulli variables.}\label{t-2}
\begin{center}
\begin{small}
\begin{tabular}[h]{r @{.} l c c c c c c c c c r @{.} l c c c c c c}
\hline
\hline
\multicolumn{3}{c}{CORR} & & \multicolumn{2}{c}{PROP} & & \multicolumn{2}{c}{MEDIAN DEV} & & \multicolumn{3}{c}{CORR} & & \multicolumn{2}{c}{PROP} & & \multicolumn{2}{c}{MEDIAN DEV}\\ \cline{1-3} \cline{5-6} \cline{8-9} \cline{11-13} \cline{15-16} \cline{18-19}
\multicolumn{2}{c}{$\rho$ W.} & $\rho$ B. & & CL-BIC & BIC & & CL-BIC & BIC & & $\rho$ W. & \multicolumn{2}{c}{$\rho$ B.} & & CL-BIC & BIC & & CL-BIC & BIC \\
\hline
 0&10            & \multirow{3}{*}{\emph{Ind}} & & 1.00 & 0.64 & & 0.0(0.0) & 2.0(0.7) & & \multirow{3}{*}{0.10 \emph{Eq}} & 0&40 & & 1.00 & 0.59 & & 0.0(0.0) & 1.0(1.5)\\
 0&15 \emph{Eq}  & & & 0.98 & 0.36 & & 1.0(0.7) & 2.0(1.5) & & & 0&50 \emph{Dec} & & 1.00 & 0.54 & & 0.0(0.0) & 2.0(0.7)\\
 0&20            & & & 0.80 & 0.08 & & 1.0(0.7) & 3.0(2.2) & & & 0&60 & & 1.00 & 0.53 & & 0.0(0.0) & 2.0(1.1) \\ \cline{1-3} \cline{11-13}
 0&40            & \multirow{3}{*}{\emph{Ind}} & & 1.00 & 0.33 & & 0.0(0.0) & 2.0(0.7) & & \multirow{3}{*}{0.15 \emph{Eq}} & 0&40 & & 0.98 & 0.32 & & 1.0(0.0) & 2.0(1.5)\\
 0&50 \emph{Dec} & & & 1.00 & 0.29 & & 0.0(0.0) & 2.0(1.5) & & & 0&50 \emph{Dec} & & 0.97 & 0.30 & & 1.0(0.4) & 3.0(1.5)\\
 0&60            & & & 1.00 & 0.14 & & 0.0(0.0) & 2.0(1.5) & & & 0&60 & & 0.95 & 0.25 & & 1.0(0.0) & 3.0(1.5)\\
\hline
\end{tabular}
\end{small}
\end{center}
\end{table*}

\textit{Simulation 1: Correlation among the edges within and between communities is introduced simultaneously throughout all blocks in the network, and not proceeding in a block-by-block fashion. Concretely, for each node $i$, all edges $\{A_{ij}\}_{i<j}$ are generated by thresholding a correlated $(N-i)$-dimensional Gaussian random vector with correlation matrix $\boldsymbol{\Sigma} = (\rho_{jl})$. Thus, in this scenario, all edges $A_{ij}$ and $A_{il}$ with common endpoint $i$ are correlated, regardless of whether $j$ and $l$ belong to the same community or not.  Cases $\rho_{jl} = \rho$ and $\rho_{jl} = \rho^{|j-l|}$, with several choices of $\rho$ are conducted. We consider a 4-community network, $\btheta = (\theta_{ab}; 1\le a \le b \le 4)'$, where $\theta_{aa}=0.35$ for all $a=1,\ldots,4$ and $\theta_{ab}=0.05$ for $1\le a < b \le 4$. Community sizes are 60, 90, 120 and 150, respectively. Results are collected in Table \ref{t-1}.}

\setcounter{dbltopnumber}{2}

\setlength{\tabcolsep}{2.9pt}

\begin{table*}[t]
\caption{\small Comparison of CL-BIC and VB over $200$ repetitions from Simulation 3. For simplicity, we omit the correlation between the corresponding Bernoulli variables.}\label{t-3}
\begin{center}
\begin{small}
\begin{tabular}[h]{c @{\;\;} c c c c c c c c c c @{\;\;} c c c c c c c c}
\hline
\hline
\multicolumn{3}{c}{CORR} & & \multicolumn{2}{c}{PROP} & & \multicolumn{2}{c}{MEDIAN DEV} & & \multicolumn{3}{c}{CORR} & & \multicolumn{2}{c}{PROP} & & \multicolumn{2}{c}{MEDIAN DEV}\\ \cline{1-3} \cline{5-6} \cline{8-9} \cline{11-13} \cline{15-16} \cline{18-19}
\multicolumn{2}{c}{$\rho$ W.} & $\rho $ B. & & CL-BIC & VB & & CL-BIC & VB & & \multicolumn{2}{c}{$\rho$ W.} & $\rho $ B. & & CL-BIC & VB & & CL-BIC & VB \\
\hline
 0.00 & \multirow{4}{*}{\emph{Eq}} & \multirow{4}{*}{\emph{Ind}} & & 1.00 & 1.00 & & 0.0(0.0) & 0.0(0.0) & & 0.00 & \multirow{4}{*}{\emph{Dec}} & \multirow{4}{*}{\emph{Ind}} & & 1.00 & 1.00 & & 0.0(0.0) & 0.0(0.0)\\
 0.10 & & & & 0.96 & 0.00 & & 1.0(0.0) & 2.0(0.0) & & 0.40 & & & & 1.00 & 1.00 & & 0.0(0.0) & 0.0(0.0)\\
 0.15 & & & & 0.88 & 0.00 & & 1.0(0.0) & 4.0(2.2) & & 0.50 & & & & 1.00 & 0.94 & & 0.0(0.0) & 1.0(0.0)\\
 0.20 & & & & 0.85 & 0.00 & & 1.0(0.0) & 5.0(1.5) & & 0.60 & & & & 1.00 & 0.56 & & 0.0(0.0) & 1.0(0.0)\\
\hline
\end{tabular}
\end{small}
\end{center}
\end{table*}

\setlength{\tabcolsep}{2.9pt}

\begin{table*}[t]
\caption{\small Comparison of CL-BIC and BIC over $200$ repetitions from Simulation 4. Before being scaled by the constant $\boldsymbol{\gamma}_n$, we selected $\btheta = (\theta_{ab}; 1\le a \le b \le 4)'$, where $\theta_{aa}=7$ for all $a=1,\ldots,4$ and $\theta_{ab}=1$ for $1\le a < b \le 4$.}\label{t-4}
\begin{center}
\begin{small}
\begin{tabular}[h]{c c c c c c c c c c c c c c c r @{.} l c}
\hline
\hline
CORR & $\gamma_n$ & & \multicolumn{2}{c}{PROP} & & \multicolumn{2}{c}{MEDIAN DEV} & & CORR & $\gamma_n$ & & \multicolumn{2}{c}{PROP} & & \multicolumn{3}{c}{MEDIAN DEV} \\ \cline{4-5} \cline{7-8} \cline{13-14} \cline{16-18}
$\rho$ MVN & & & CL-BIC & BIC & & CL-BIC & BIC & & $\rho$ MVN & & & CL-BIC & BIC & & \multicolumn{2}{c}{CL-BIC} & BIC \\
\hline
 \multirow{2}{*}{0.20 \emph{Eq}} & 0.02 & & 0.84 & 0.35 & & 1.0(1.5) & 2.0(1.5) & & \multirow{2}{*}{0.60 \emph{Dec}} & 0.02 & & 0.92 & 0.52 & & -1&0(1.5) & 2.0(1.5)\\
 & 0.03 & & 0.96 & 0.58 & & 1.0(0.0) & 3.0(1.5) & & & 0.03 & & 1.00 & 0.81 & & 0&0(0.0) & 1.0(1.5)\\ \cline{1-2} \cline{10-11}
 \multirow{2}{*}{0.30 \emph{Eq}} & 0.02 & & 0.70 & 0.31 & & 1.0(0.4) & 2.0(0.7) & & \multirow{2}{*}{0.70 \emph{Dec}} & 0.02 & & 0.83 & 0.41 & & -1&0(1.5) & 2.0(1.5)\\
 & 0.03 & & 0.93 & 0.52 & & 1.0(0.6) & 3.0(1.5) & & & 0.03 & & 1.00 & 0.77 & & 0&0(0.0) & 1.0(0.7)\\ \cline{1-2} \cline{10-11}
 \multirow{2}{*}{0.40 \emph{Eq}} & 0.02 & & 0.43 & 0.21 & & 1.0(1.5) & 2.0(1.5) & & \multirow{2}{*}{0.80 \emph{Dec}} & 0.02 & & 0.69 & 0.22 & & -1&0(1.5) & 3.0(2.4)\\
 & 0.03 & & 0.85 & 0.51 & & 1.0(1.9) & 3.0(1.9) & & & 0.03 & & 0.98 & 0.69 & & -1&0(0.4) & 1.0(1.5)\\
\hline
\end{tabular}
\end{small}
\end{center}
\end{table*}

\textit{Simulation 2: Correlation among the edges within ($\rho$ W.) and between ($\rho$ B.) communities is introduced block-wisely. Concretely, for each node $i$, all edges $A_{ij}$ and $A_{il}$ are generated independently whenever $j$ and $l$ belong to different communities. If $j$ and $l$ belong to the same community, edges $A_{ij}$ and $A_{il}$ are generated by thresholding a correlated Gaussian random vector with correlation matrix $\boldsymbol{\Sigma} = (\rho_{jl})$. Parameter settings are identical to Simulation 1, with results collected in Table \ref{t-2}.}

\textit{Simulation 3: Correlation settings are the same as in Simulation 2, but we change the value of the parameter $\btheta$ to allow for more general network topologies. We set $\btheta = (\theta_{ab}; 1\le a \le b \le 4)'$ with $\theta_{aa}=\theta_{b4}=0.35$ for all $a=1,\ldots,4$ and $b=1,2,3$. The remaining entries of $\btheta$ are set to $0.05$. Hence, following \citet{LatoucheEtal2012}, vertices from community 4 connect with probability $0.35$ to any other vertices in the network, forming a community of only hubs. Community sizes are the same as in Simulation 1, with results collected in Table \ref{t-3}.}

\textit{Simulation 4: We follow the approach of \citet{ZhaoEtal2012} in choosing the parameters $(\btheta, \bomega)$ to generate networks from the degree-corrected blockmodel. Thus, the identifiability constraint $\sum_{i}\omega_i\bone\{z_i = a\} = 1$ for each community $1 \leq a \leq K$ is replaced by the requirement that the $\omega_i$ be independently generated from a distribution with unit expectation, fixed here to be
$$
\omega_i = \left \{ \begin{array}{l l}
                \eta_i, & \text{ w.p. $0.8$,} \\
                2/11,   & \text{ w.p. $0.1$,} \\
                20/11,  & \text{ w.p. $0.1$,}
                \end{array} \right .
$$
where $\eta_i$ is uniformly distributed on the interval $[0,2]$. The vector $\btheta$, in a slight abuse of notation, is reparametrized as $\btheta_n = \boldsymbol{\gamma}_n \btheta$, where we vary the constant $\boldsymbol{\gamma}_n$ to obtain different expected degrees of the network. Correlation settings and community sizes are the same as in Simulation 1, with results presented in Table \ref{t-4}, where choices for $\boldsymbol{\gamma}_n$ and $\btheta$ are specified.}

When the stochastic blockmodels are contaminated by the imposed correlation structure, which is expected in real-world networks, CL-BIC outperforms BIC overwhelmingly. Tables \ref{t-1}--\ref{t-2} show the improvement is more significant when the imposed correlation is larger. For instance, in the block-wise correlated case of Table \ref{t-2}, when we only have within-community correlation $\rho_{\emph{Dec}}=0.60$, CL-BIC does the right selection in all cases, while BIC is only successful in $14\%$ of 200 trials.

As shown in Table \ref{t-3} for the model with a community of only hubs, if the network is generated from a purely stochastic blockmodel, or if the contaminating correlation is not too strong, CL-BIC and VB have similar performance in selecting the correct $K=4$. But again, as the imposed correlation increases, VB fails to make the right selection more often than CL-BIC. This is particularly true in the $\rho_{\emph{Eq}}=0.20$ case, where CL-BIC makes the right selection in $85\%$ of simulated networks, whereas VB fails in all cases, yielding models with a median of $9$ communities.

The same pattern translates into the DCBM setting of Table \ref{t-4}, where smaller values of $\boldsymbol{\gamma}_n$ yield sparser networks. The community number selection problem becomes more difficult as $\boldsymbol{\gamma}_n$ decreases, as degrees for many nodes are small, yielding noisy individual effect estimates $\hat{\omega}_i = d_i/\sum_{j: z_j = z_i}d_j$. Nevertheless, the CL-BIC approach consistently selects the correct number of communities more frequently than BIC over different correlation settings.

\begin{figure*}[t]
\begin{center}
\includegraphics[width=5in, height=3.5in]{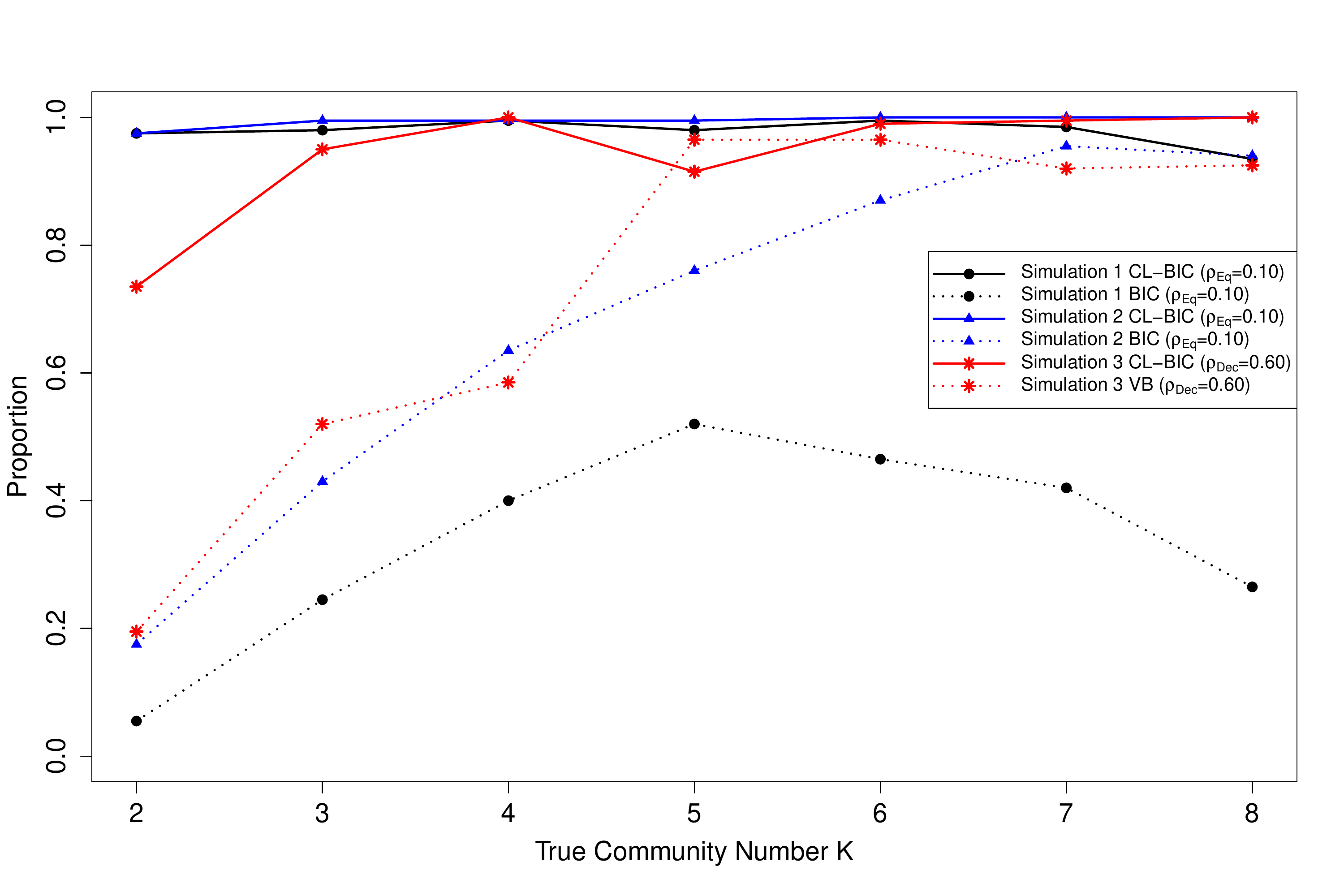}
\caption{\small (Color online) Comparisons between different methods for selecting the true community number $K$ in the standard blockmodel settings of Simulations 1 -- 3. Along the $y$-axis, we record the proportion of times the chosen number of communities for each of the different criteria for selecting $K$ agrees with the truth.}\label{DivergingK}
\end{center}
\end{figure*}

In addition, Figure \ref{DivergingK} presents simulation results where the true community number $K$ increases from $2$ to $8$. Following our previous examples, community sizes grow according to the sequence $(60,90,120,150,60,90,120,150)$. The selected correlation-contaminated stochastic blockmodels are $\rho_{Eq}=0.10$ from Simulation 1, within-community correlation $\rho_{Eq}=0.10$ from Simulation 2, and within-community correlation $\rho_{Dec}=0.60$ from Simulation 3. As $K$ increases and enough vertices are added into the network, CL-BIC tends to correctly estimate the true community number in all simulation settings. Even in this scenario with a growing number of communities, the proportion of times CL-BIC selects the true $K$ is always greater than the corresponding BIC or VB estimates.

Before moving to the last simulation example, we would like to  define two measures to quantify the accuracy of a given node label assignment. The first measure is a ``goodness-of-fit" (GF) measure defined as
\begin{equation}\label{GF}
\begin{split}
GF(\boldsymbol{z},\boldsymbol{\hat{z}_k}) & =\sum_{i<j} \left( \mathbbm{1}\{z_i=z_j\}\mathbbm{1}\{\hat{z}_i=\hat{z}_j\} + \mathbbm{1}\{z_i \neq z_j\}\mathbbm{1}\{\hat{z}_i \neq \hat{z}_j\} \right) \Big/ {N \choose 2},
\end{split}
\end{equation}
where $\boldsymbol{z}$ represents the true community labels and $\boldsymbol{\hat{z}_k}$ represents the community assignments from an estimator. Thus, the measure $GF(\boldsymbol{z},\boldsymbol{\hat{z}_k})$ calculates the proportion of pairs whose estimated assignments agree with the correct labels in terms of being assigned to the same or different communities, and is commonly known as the \emph{Rand Index} \citep{Rand1971} in the cluster analysis literature.

The second measure is motivated from the ``assortativity" notion. The ratio of the median within community edge number to that of the between community edge number (MR) is defined as
\begin{equation}\label{MR}
MR(\boldsymbol{\hat{z}_k}) = \median_{a=1,\ldots,k} \left(m_{aa}\right) \Big/ \median_{a \neq b} \left(m_{ab}\right),
\end{equation}
where $k$ is the number of communities implied by $\boldsymbol{\hat{z}_k}$ and $m_{ab}$ is the total number of edges between communities $a$ and $b$, as given by the community assignment $\boldsymbol{\hat{z}_k}$. It is clear that for both measures, a higher value indicates a better community detection performance.

As a final simulation example, we analyze the performance of CL-BIC and BIC for a growing number of communities under the degree-corrected blockmodel. While the reparametrized vector $\btheta_n = \boldsymbol{\gamma}_n \btheta$ remains as in Simulation 4, the $\omega_i$ are now independently generated from Uniform$(1/5,9/5)$. The results are collected in Table \ref{t-5}, where we also record the performance of the SCORE algorithm under the true $K$, along with the goodness-of-fit (GF) and median ratio (MR) performance measures introduced in \eqref{GF} and \eqref{MR}, respectively.

\setlength{\tabcolsep}{4.4pt}

\setcounter{dbltopnumber}{2}

\begin{table*}[t]
\small
\begin{threeparttable}
\caption{\small Comparison of CL-BIC and BIC over $200$ repetitions from the DCBM case in Simulation 4, with $(\rho_{Eq}=0.2,\boldsymbol{\gamma}_n=0.03)$, where the individual effect parameters $\omega_i$ are now generated from a Uniform$(1/5,9/5)$ distribution.}\label{t-5}
\begin{center}
\begin{small}
\begin{tabular}[h]{c c c c c c c r @{} l c c c c c c c c}
\hline
\hline
 & \multicolumn{3}{c}{SCORE Performance} & & \multicolumn{6}{c}{CL-BIC} & & \multicolumn{5}{c}{BIC} \\ \cline{2-4} \cline{6-11} \cline{13-17}
$K$ & Misc. R. & Orac. Err. & Est. Err. & & PROP & MD & \multicolumn{2}{c}{RSD} & GF & MR & & PROP & MD & RSD & GF & MR \\
\hline
2 & 0.02 & 0.51 & 0.54 & & 0.88 & 1 & 0&    & 0.96 & 7.33 & & 0.10 & 2 & 0.75 & 0.73 & 3.37 \\
3 & 0.03 & 0.53 & 0.55 & & 0.93 & 1 & 0&.75 & 0.97 & 7.87 & & 0.09 & 3 & 1.49 & 0.88 & 3.91 \\
4 & 0.03 & 0.55 & 0.58 & & 0.86 & 1 & 0&    & 0.97 & 8.11 & & 0.16 & 3 & 1.49 & 0.91 & 5.41 \\
5 & 0.04 & 0.58 & 0.62 & & 0.56 & 1 & 0&    & 0.96 & 6.35 & & 0.09 & 3 & 2.05 & 0.92 & 5.92 \\
6 & 0.05 & 0.60 & 0.64 & & 0.47 & 1 & 1&.49 & 0.96 & 7.10 & & 0.09 & 2 & 2.24 & 0.94 & 7.08 \\
7 & 0.05 & 0.63 & 0.66 & & 0.39 & 1 & 1&.49 & 0.97 & 6.77 & & 0.09 & 3 & 1.49 & 0.95 & 6.73 \\
8 & 0.08 & 0.63 & 0.66 & & 0.29 & 1 & 1&.49 & 0.97 & 7.24 & & 0.02 & 3 & 2.24 & 0.96 & 6.80 \\
\hline
\end{tabular}
\begin{tablenotes}[para,flushleft]
NOTE: PROP, proportion; MD, median deviation; RSD, robust standard deviation; GF, goodness-of-fit measure; MR: median ratio measure. Misc. R. denotes the misclustering rate of the SCORE algorithm. For $\Omega=E_{\btheta}(\bA)$, Orac. Err. and Est. Err. are  $\norm{\Omega_{O}-\Omega}/\norm{\Omega}$ and $\norm{\Omega_{SC}-\Omega}/\norm{\Omega}$, respectively, where $\norm{\cdot}$ denotes Frobenius norm. Here, $\Omega_{O}$ denotes the estimate of $\Omega$ under the \textit{oracle} scenario where we know the true community assignment $\boldsymbol{z} \in \{1, \ldots, K\}^N$, and $\Omega_{SC}$ is the estimate of $\Omega$ using the SCORE labeling vector.
\end{tablenotes}
\end{small}
\end{center}
\end{threeparttable}
\vspace{-0.25in}
\end{table*}

The true community number and community sizes grow as in the case for the standard blockmodel described in Figure \ref{DivergingK}. Although CL-BIC performs uniformly better than BIC across all validating criteria and throughout all $K$, the procedure does not appear to yield model selection consistent results in this example. Aside from the fact that the introduced correlation is not exponentially decaying, this poor performance as $K$ increases can also be explained by the difficulty in estimating the DCBM parameters $(\btheta, \bomega)$ in a scenario where several vertices have potentially low degrees. Indeed, even in the \textit{oracle} scenario where we know the true community labels $z_i$ ahead of time, and for a relatively small misclustering rate of the SCORE algorithm, Table \ref{t-5} exhibits the difficulty in obtaining accurate estimates $(\boldsymbol{\hat{\theta}}_{\text{C}},\boldsymbol{\hat{\omega}}_{\text{C}})$, and in evaluating the CL-BIC criterion functions \eqref{criterion}, under this increasing $K$ scenario for the DCBM. Whether the increased number of parameters in the DCBM has an effect on the consistency results of CL-BIC as $K$ increases is also an interesting line of future work.

\begin{figure*}[t]
\begin{center}
\subfigure[$k=3$ (CL-BIC-SBM)]{
\includegraphics[width=2.099in, height=2.37in]{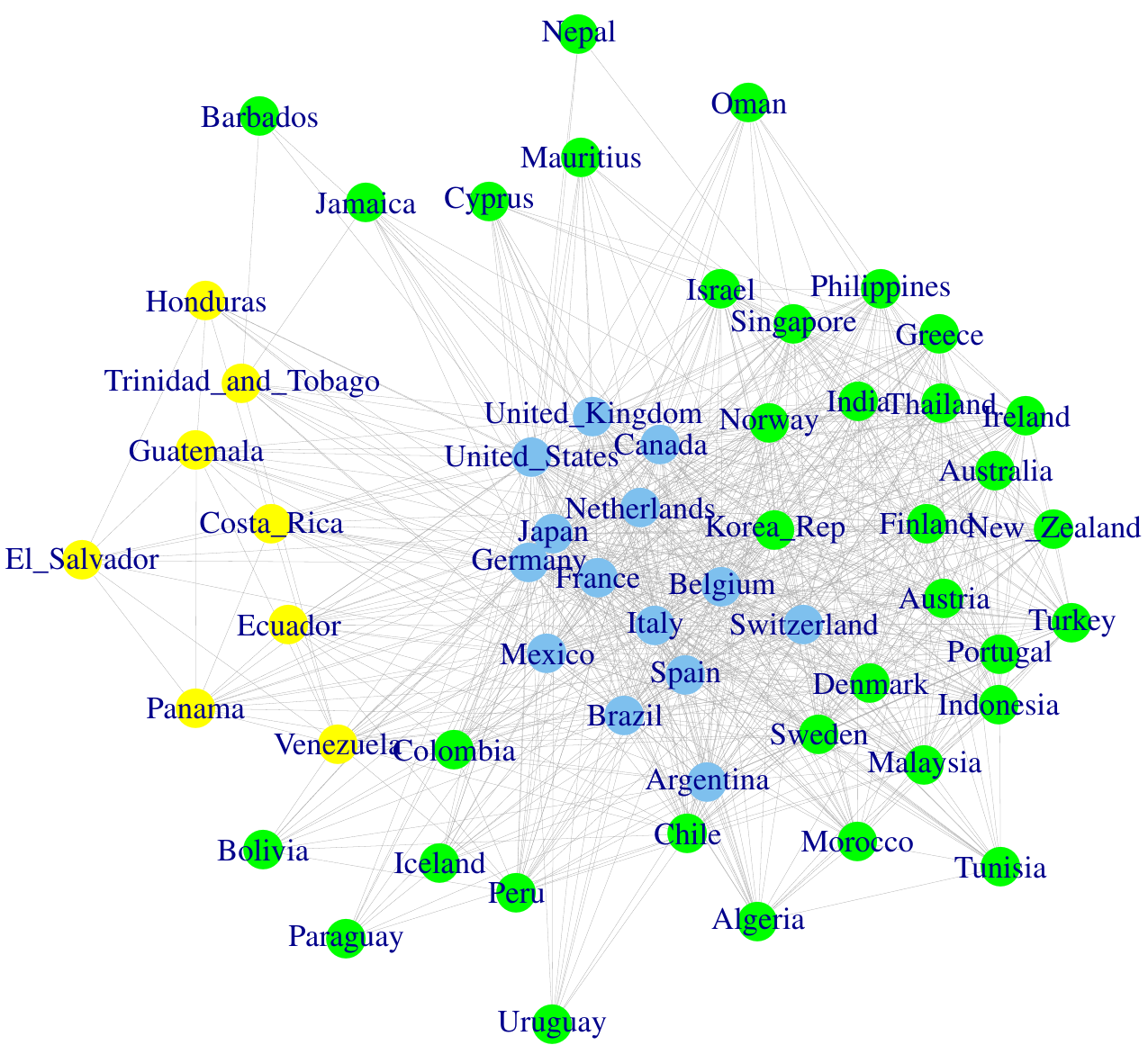}}
\subfigure[$k=7$ (VB-SBM)]{
\includegraphics[width=2.1in, height=2.37in]{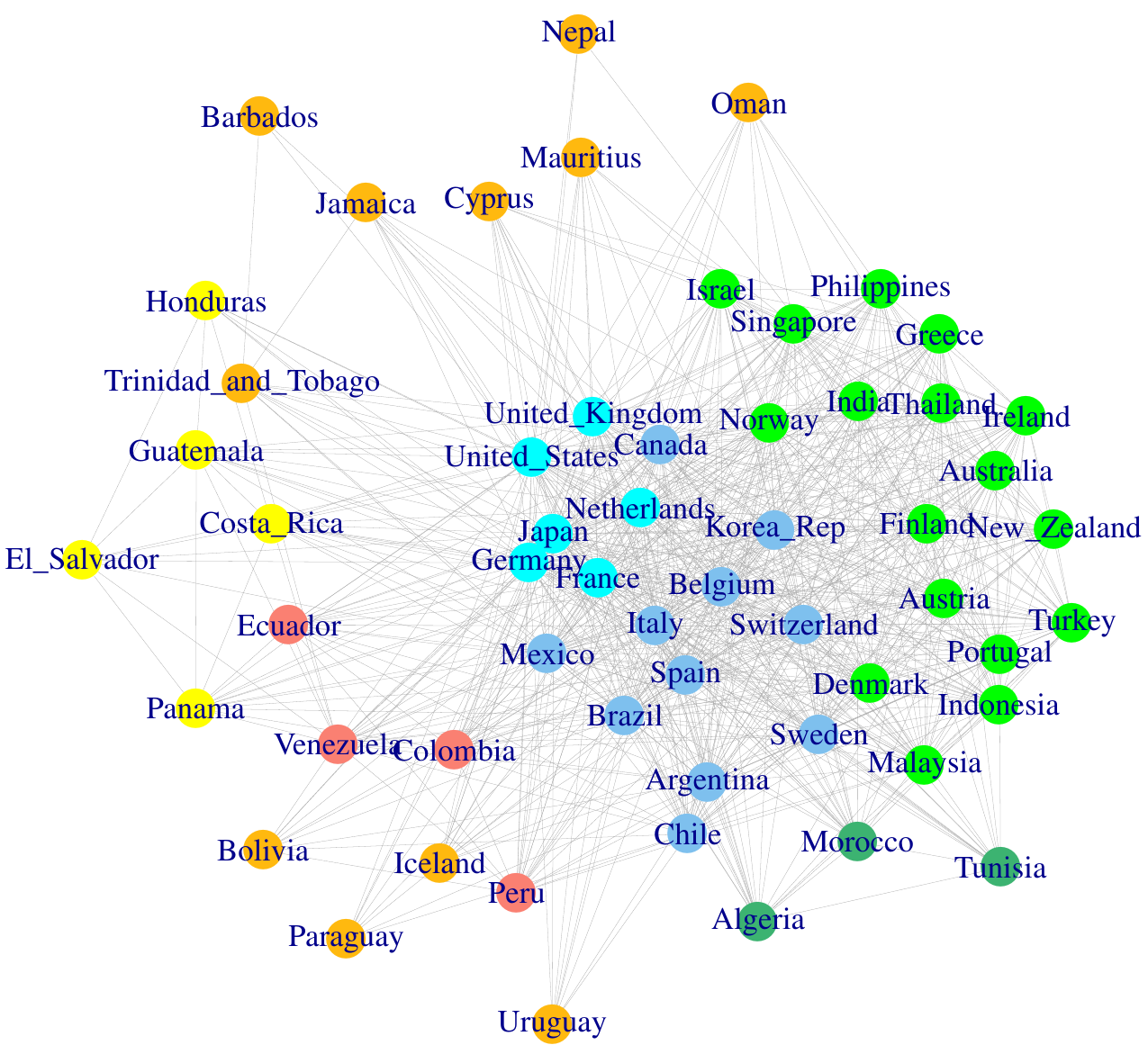}}
\subfigure[$k=10$ (BIC-SBM)]{
\includegraphics[width=2.099in, height=2.37in]{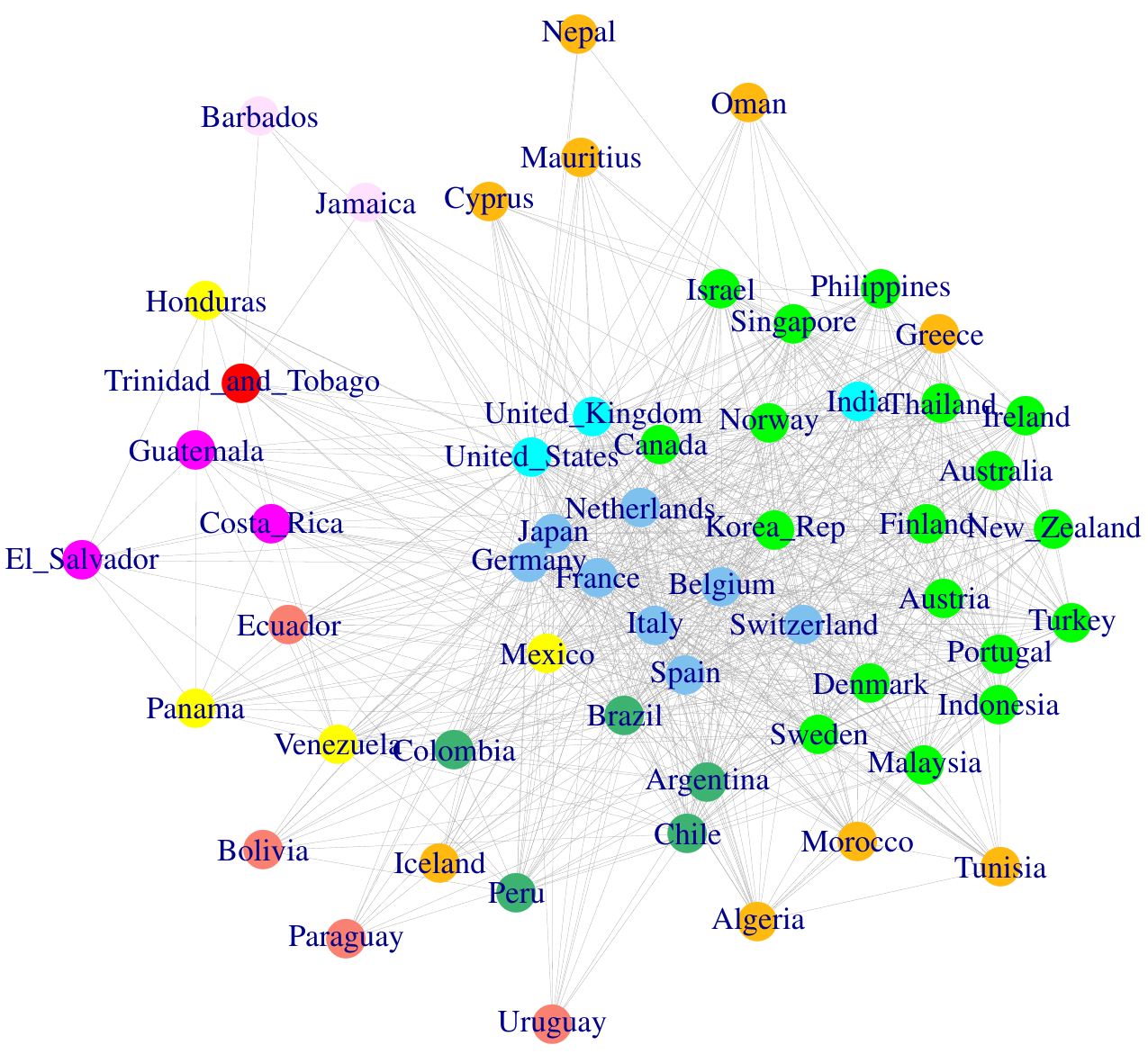}}
\caption{\small (Color online) Largest connected component of the international trade network for the year $1995$.}\label{Trade}
\end{center}
\end{figure*}

\subsection{Real Data Analysis}\label{realdata}

\subsubsection{International Trade Networks}

We first study an international trade dataset originally analyzed in \citet{WestveldHoff2011}, containing yearly international trade data between $N=58$ countries from $1981-2000$. For a more detailed description of this dataset, we refer the interested reader to the Appendix in \citet{WestveldHoff2011}. In our numerical comparisons between CL-BIC and BIC paired with the standard stochastic blockmodel log-likelihood \eqref{eq-lik}, we focus on data from year 1995. For this network, an adjacency matrix $\bA$ can be formed by first considering a weight matrix $\bW$ with $W_{ij} = \text{Trade}_{i,j} + \text{Trade}_{j,i}$, where $\text{Trade}_{i,j}$ denotes the value of exports from country $i$ to country $j$. Finally, we define $A_{ij}=1$ if $W_{ij} \geq W_{\alpha}$, and $A_{ij}=0$ otherwise; here $W_{\alpha}$ denotes the $\alpha$\emph{-th} quantile of $\{W_{ij}\}_{1 \leq i < j \leq N}$. For the choice of $\alpha = 0.5$, Figure \ref{Trade} shows the largest connected component of the resulting network. Panel (a) shows CL-BIC selecting $3$ communities, corresponding to countries with the highest GDPs (dark blue), industrialized European and Asian countries with medium-level GDPs (green), and developing countries in South America with the smallest GDPs (yellow). Next, in panel (b) we also show the variational Bayes solution corresponding to $k=7$, providing finer communities for some Central and South American neighboring countries (yellow and pink, respectively) but fragmenting the high- and medium-level GDP countries into ambiguous communities. For instance, it is not clear why countries like Bolivia and Nepal belong to the same community (orange) or why the Netherlands, rather than Brazil or Italy, joined the community of countries with the highest GDPs (light blue). At last, panel (c) corresponds to the final BIC model selecting $10$ communities. Under this partition, South American countries are now split into $6$ ``noisy" communities, while high GDP countries are unnecessarily fragmented into two.

We believe CL-BIC provides a better model than traditional BIC, yielding communities with countries sharing similar GDP values without dividing an entire continent into $6$ smaller communities. On the contrary, BIC selects a model containing communities of size as small as one, which are of little, if any, practical use. The variational Bayes approach provides a meaningful solution in this example, exhibiting a similar performance as in \cite{LatoucheEtal2012} in terms of providing some finer community assignments.

\begin{figure*}[t]
\begin{center}
\subfigure[$k=6$ (CL-BIC-DCBM)]{
\includegraphics[width=2.095in, height=1.8in]{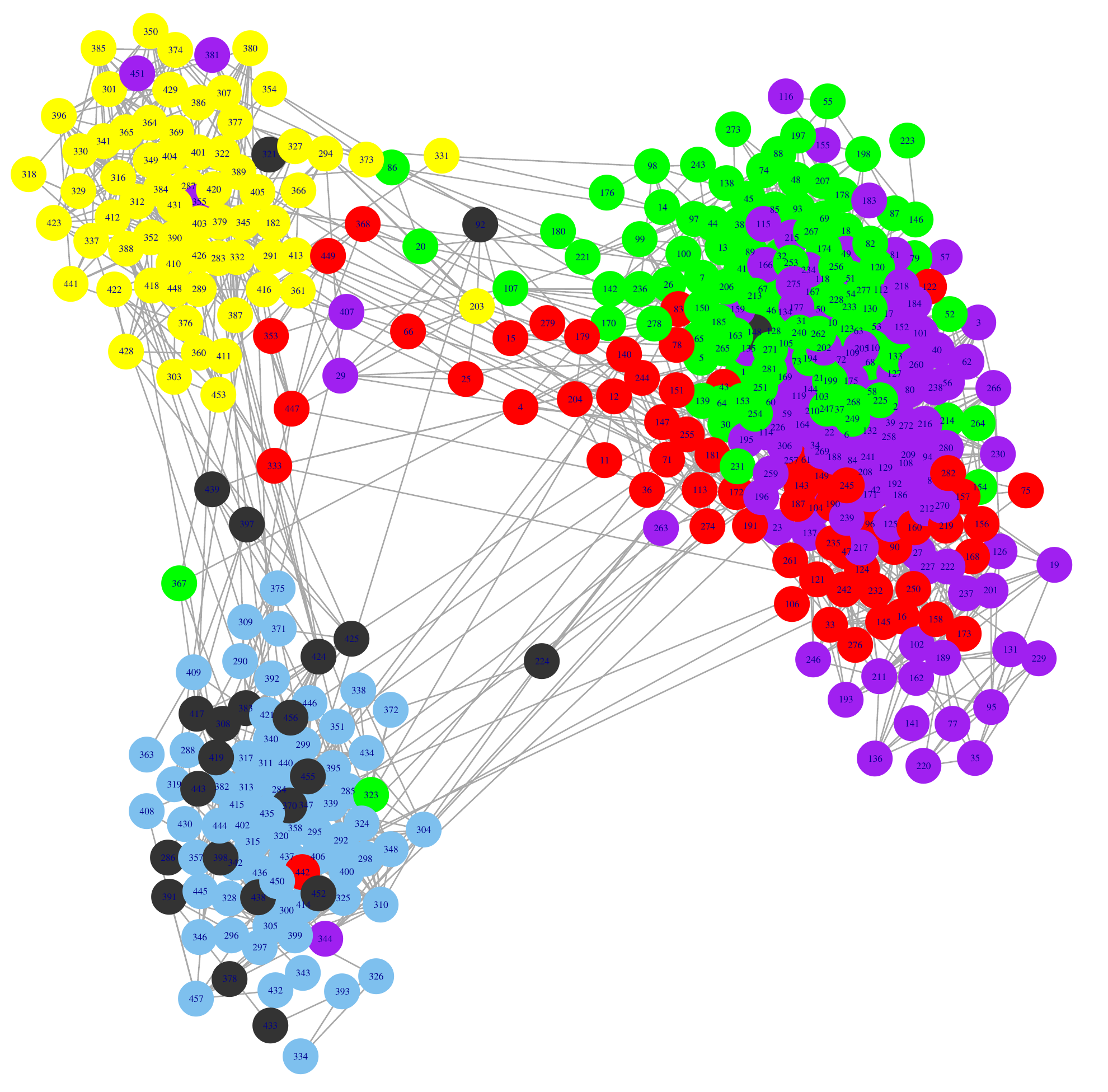}}
\subfigure[$k=9$ (BIC-DCBM)]{
\includegraphics[width=2.095in, height=1.8in]{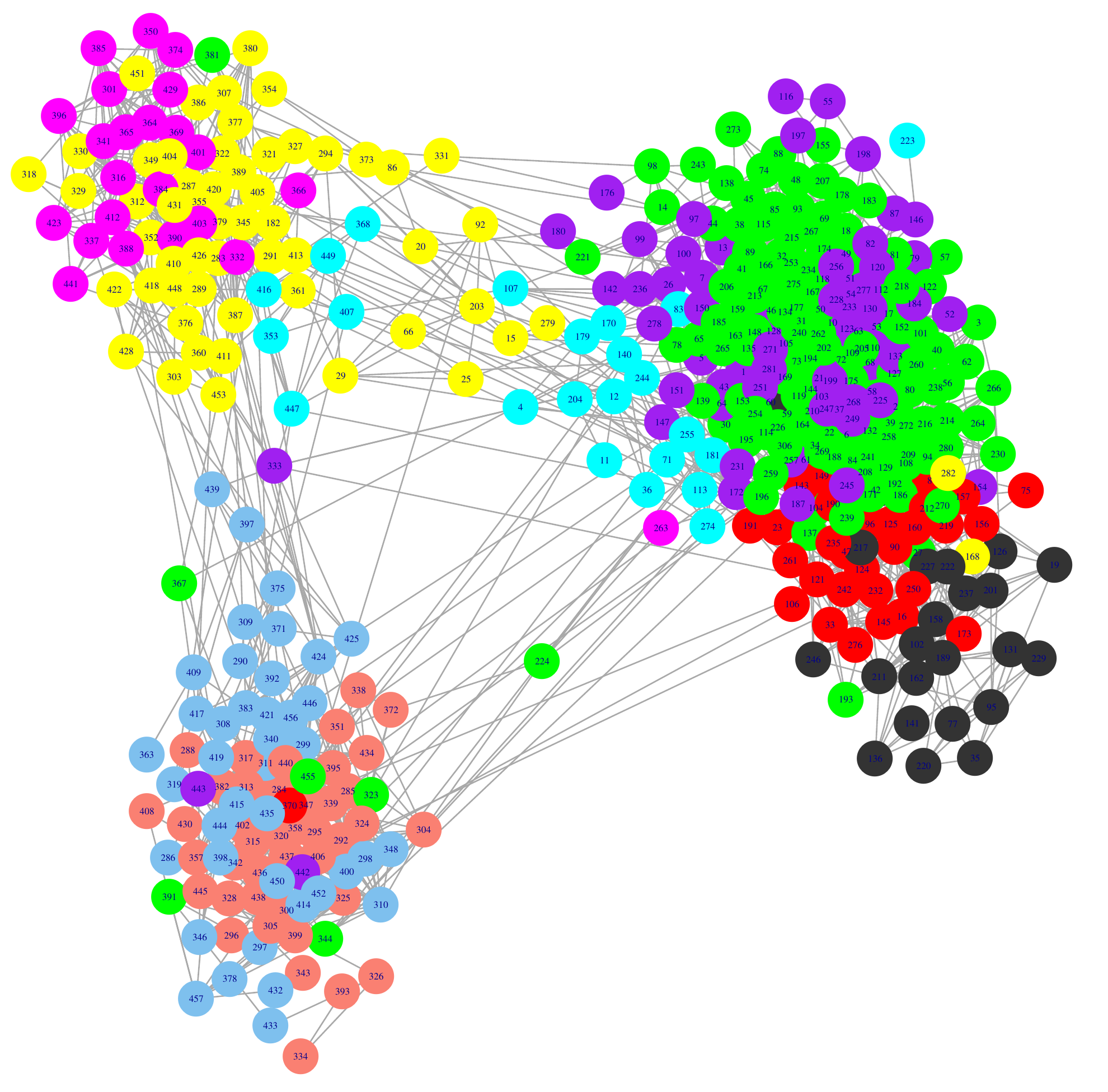}}
\subfigure[``Truth" $(K=6)$]{
\includegraphics[width=2.095in, height=1.8in]{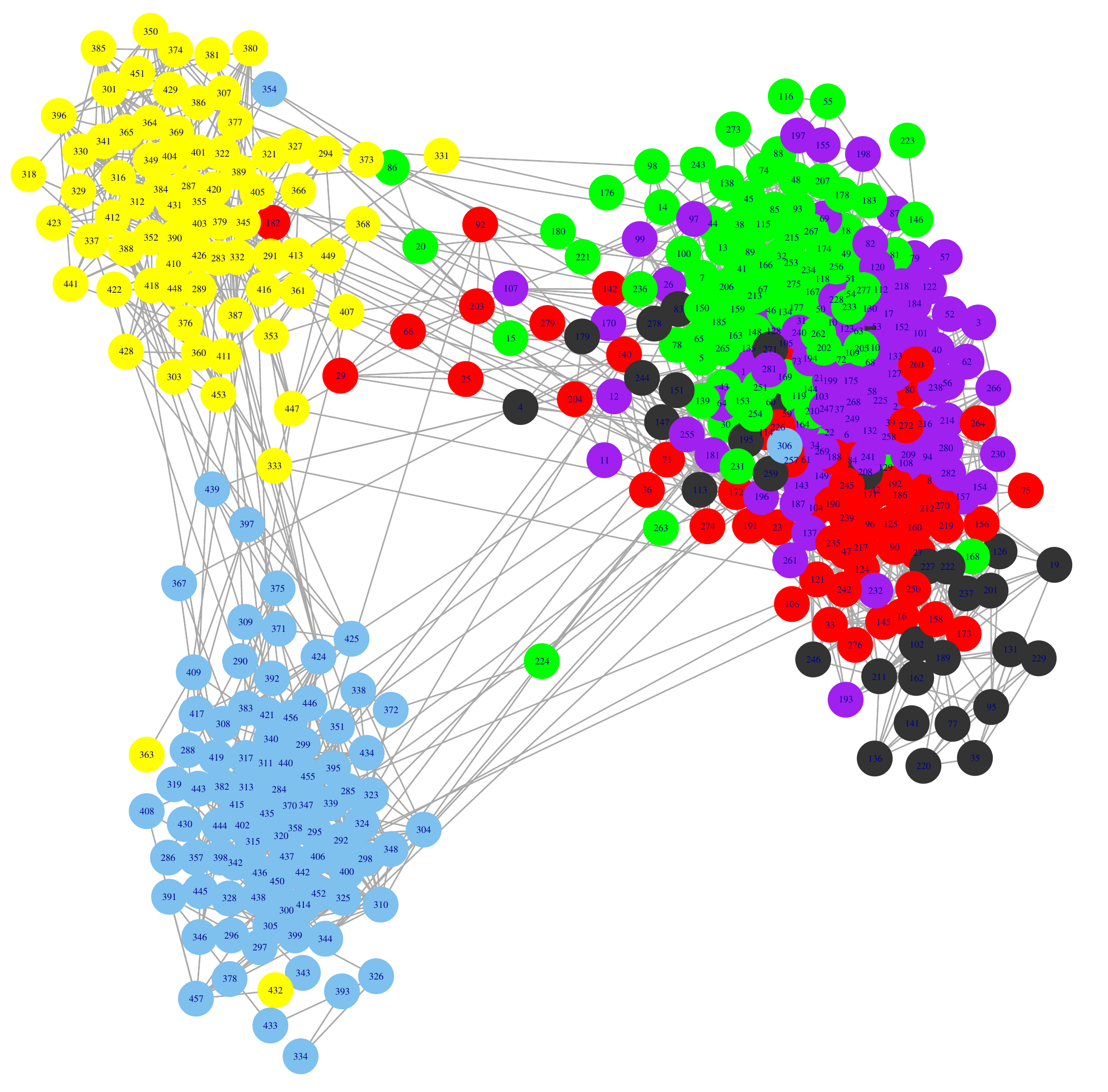}}
\caption{\small (Color online) Largest connected component of the school friendship network. Panel (c) shows the ``true" grade community labels: $7$th (blue), $8$th (yellow), $9$th (green), $10$th (purple), $11$th (red), and $12$th (black).}\label{S7}
\end{center}
\end{figure*}

\subsubsection{School Friendship Networks}

Now, we consider a school friendship network obtained from the National Longitudinal Study of Adolescent Health (\href{http://www.cpc.unc.edu/projects/addhealth}{http://www.cpc.unc.edu/projects/addhealth}). For this network, $A_{ij}=1$ if either student $i$ or $j$ reported a close friendship tie between the two, and $A_{ij}=0$ otherwise. We  focus on the network of school 7 from this dataset, and our comparisons between CL-BIC and BIC are done with respect to the degree-corrected blockmodel log-likelihood \eqref{eq-lik-dcbm}. With $433$ vertices, Figure \ref{S7} shows the largest connected component of the resulting network. As shown in panel (a), CL-BIC selects the true community number $K=6$, roughly agreeing with the actual grade labels, except for the black community. BIC, shown in panel (b), selects $9$ communities, unnecessarily splitting the $7$th and $8$th graders. The ``true" friendship network is shown in panel (c).

We still conclude CL-BIC performs better than traditional BIC. Except for the misallocation of the black community of $12$th graders, the model selected by CL-BIC correctly labels most of the remaining network. While BIC partially separates the 10th graders and the 12th graders, a substantial portion of the 10th graders are absorbed into the 9th grader community (green). In addition, BIC further fragments $7$th and $8$th graders into ``noisy" communities. This is an extremely difficult community detection problem since, even for a ``correctly" specified $k=6$, SCORE fails to assign all $12$th graders to their corresponding true grade. The black community selected by SCORE in panel (a) mainly corresponds to female students and hispanic males, reflecting perhaps closer friendship ties among a subgroup of students recently starting junior high school.

Using the ``goodness-of-fit" measure defined in \eqref{GF}, we found that the CL-BIC community assignment leads to $GF(\boldsymbol{z},\boldsymbol{\hat{z}_6}) = 0.811$, which is slightly better than the $GF(\boldsymbol{z},\boldsymbol{\hat{z}_9}) = 0.810$ obtained for BIC. For the MR measure given in \eqref{MR}, the results for CL-BIC and BIC are $MR(\boldsymbol{\hat{z}_6})=40.8$ and $MR(\boldsymbol{\hat{z}_9})=33.3$, respectively, again indicating the superiority of the CL-BIC solution paired with SCORE.

In both examples, BIC tends to overestimate the ``true" community number $K$, rendering very small communities which are in turn penalized under the CL-BIC approach. This means CL-BIC successfully remedies the robustness issues brought in by spectral clustering, due to the misspecification of the underlying stochastic blockmodels, and effectively captures the loss of variance produced by using traditional BIC.

\section{Discussion}\label{discussion}

There has been a tremendous amount of research in recovering the underlying structures of network data, especially on the community detection problem. Most of the existing work has focused on studying the properties of the stochastic blockmodel and its variants without looking at the possible model misspecification problem. In this paper, under the standard stochastic blockmodel and its variants, we advocate the use of composite likelihood BIC for selecting the number of communities due to its simplicity in implementation and its robustness against correlated binary data.

Some extensions are possible. For instance, the proposed methodology in this work is based on the spectral clustering and SCORE algorithms, and it would be interesting to explore the combination of the CL-BIC with other community detection methods. In addition, most examples considered here are dense graphs, which are common but cannot exhaust all scenarios in real applications. Another open problem is to study whether the CL-BIC approach is consistent for the degree-corrected stochastic blockmodel, which is not necessarily true from our numerical studies.

\section*{Supplementary Materials}

\begin{description}

\item[R Code and Trade Dataset:] The R codes can be used to replicate the simulation studies and the real data analysis. The international trade network dataset is also included. More details can be found in the file README contained in the zip file (CLBIC.zip).

\end{description}

\section*{Acknowledgements}
The authors thank the editor, the associate editor, and two anonymous referees for their constructive comments which have greatly improved the paper. Yu is partially supported by Richard Samworth's Engineering and Physical Sciences Research Council Early Career Fellowship EP/J017213/1. Feng is partially supported by NSF grant DMS-1308566.

\bibliography{network}
\bibliographystyle{apa}

\end{document}